# Anisotropic Behavior of Excitons in Single Crystal α-SnS

Van Long Le[1,3†], Do Duc Cuong[2†], Hoang Tung Nguyen[1,3], Xuan Au Nguyen[1], Bogyu Kim[1], Kyujin Kim[1], Wonjun Lee[1], Soon Cheol Hong[2*], Tae Jung Kim[1,4*], and Young Dong Kim[1*]

[1]*Department of Physics, Kyung Hee University, Seoul 02447, Republic of Korea*

[2]*Department of Physics and Energy Harvest-Storage Research Center, University of Ulsan, Ulsan 44610, Republic of Korea*

[3]*Institute of Materials Science, Vietnam Academy of Science and Technology, Hanoi 100000, Vietnam*

[4]*Center for Converging Humanities, Kyung Hee University, Seoul 02447, Republic of Korea*

[†] Authors contributed equally

Corresponding authors:

Y. D. Kim (email: ydkim@khu.ac.kr), T. J. Kim (email: tjkim@khu.ac.kr), S. C. Hong (email: schong@ulsan.ac.kr)

We investigate analytically the anisotropic dielectric properties of single crystal α-SnS near the fundamental absorption edge by considering atomic orbitals. Most striking is the excitonic feature in the armchair- (*b*-) axis direction, which is particularly prominent at low temperatures. To determine the origin of this anisotropy, we perform first-principles calculations using the GW0 Bethe-Salpeter equation (BSE) including the electron-hole interaction. The results show that the anisotropic dielectric characteristics are a direct result of the natural anisotropy of *p* orbitals. In particular, this dominant excitonic feature originates from the $p_y$ orbital at the saddle point in the Γ-*Y* region.

# I. INTRODUCTION

SnS and related materials have received much attention because of potential applications in thermoelectric device [1–5], thin-film photovoltaic, [6–10], and photoelectronic materials [11–16] technologies. Accordingly, many studies on physical, electrical, and optical properties of SnS have been reported on both bulk [4,17–23] and lower dimensional structures [15,24–27]. The anisotropic characteristics of its optical properties are well known as a result of data obtained by electron-energy-loss spectroscopic [21], core-level spectroscopic [20], infrared reflectance [17], and spectroscopic ellipsometry (SE) measurements [18,23].

SnS crystallizes in an orthorhombic structure belonging to the space group Pnma. Figure 1(a) shows the atomic structure, which consists of valence bonds between each Sn atom (dark grey) and three adjacent S (bright gray) atoms to form a puckered honeycomb network, resulting in unusually strong in-plane anisotropy. In the $c$ direction, the SnS layers are well separated by weak van der Waals bonding. This structure is therefore an intermediate between a two-dimensional (2D) and three-dimensional (3D) materials. The Brillouin zone (BZ) of the Pnma structure, shown in Fig. 1(b), clearly demonstrates this anisotropy.

Previous theoretical calculation reveals that this anisotropy results in two band gaps at 1.21 eV and 1.54 eV along the zigzag- ($a$-) and $b$-directions, respectively [19]. Makinistian and Albanesi explained this anisotropy based on partial density of states [19]. Based on room-temperature pseudodielectric function data obtained by SE, Banai *et al*. evaluated critical points (CPs) of band-to-band transitions [23]. Recently, some of us reported dielectric-tensor data of SnS from energies from 0.74 to 6.42 eV and temperatures from 27 to 350 K [18]. The exciton feature is prominent in the armchair direction, but is absent in the other directions.

Here, we employ first principle calculations to understand the origin of the dielectric anisotropy of single-crystal SnS on the atomic scale. We find that the GW0 Bethe-Salpeter equation (BSE) with the electron-hole interaction included accurately describes the data. The excitonic feature is due to band-to-band transitions at a saddle point in the $\Gamma - Y$ region, and is formed mainly from $s$ and $p_y$ orbitals. The direct band gap near 1.4 eV occurs in $a$-direction, and is found to be due to band-to-band transitions in the $\Gamma - X$ involving $s$ and $p_x$ orbitals.

## II. BASICS

The data analyzed here were obtained by SE on single-crystal SnS grown at 960 ºC by the temperature-gradient method. Details of the growth and structural characterization are given in Ref. 28. Measurements were done with a rotating-compensator SE (J.A. Woollam Inc., RC2 model). Data for all three components of the dielectric tensor were obtained from 0.74 to 6.42 eV with the sample at 27 K.

First-principles density-functional-theory (DFT) calculations were performed within the framework of the projector-wave formalism [29], as implemented in the Vienna *ab initio* simulation package (VASP) [30]. The structural-parameter values of SnS were those determined experimentally and reported in Ref. 22. The generalized-gradient-approximation (GGA) with Perdew, Burke, and Ernzerhof (PBE) parameterization was used to describe the exchange correlation functional [31,32]. The plane-wave cutoff for kinetic energy was 350 eV. BZ integrations were done with a $8\times8\times4$ $\Gamma$–centered *k*-point grid.

Excitonic properties were determined by solving the Bethe-Salpeter equation (BSE) with the Tamm-Dancoff approximation on the GW0 quasiparticle band structure [33]. Indirect transitions are not included in present study, since we are dealing with optical transitions only. The six highest valence bands and the eight lowest conduction bands were included as bases for the excitonic state to assure convergence to the lowest electron-hole excitation energy, which determines the excitonic binding energy.

## III. RESULTS AND DISCUSSION

Figure 1(c) shows the anisotropic behavior of the pseudodielectric function of SnS according to different azimuthal angles ($\theta$) at an angle of incidence (AOI) of 70º at room temperature. It is clear that optical absorption along the *a*- and *b*-directions are significantly different. Here, $\theta = 0º$ corresponds to the zigzag-direction. These data clearly show the highly anisotropic properties of SnS along the two principal axes, as seen in the band gaps. These are about 1.30 and 1.60 eV along the zigzag- and armchair-directions, respectively, in good agreement with previous reports [19,23]. As shown in Fig. 1(d), the full rotation of the azimuth angle confirms the 2-fold rotational symmetry of the orthorhombic structure of single-crystal SnS.

Figure 2 shows the dielectric tensor of single crystal SnS at 27 and 300 K. Figures 2(a-b) show the real and imaginary parts on the cleavage plane at 27 K, in which the dashed, solid, and dashed-dotted curves represent dielectric functions at azimuthal angles $\theta = 0, 45,$ and $90°$, respectively. It is evident that excitonic feature essentially disappears in the zigzag-direction. At $\theta = 45°$, the dielectric function has both contributions, exhibiting a fundamental bandgap of about 1.4 eV and an excitonic peak near 1.7 eV. Similarly, Figs. 2(c-d) show the dielectric functions at different azimuthal angle at 300 K. Here, the exciton completely vanishes in the armchair direction. Figure 2 seems to indicate that the CPs along the principal axes may have the same origins in the band structure. We note that the excitonic feature is also observed along the armchair axis of GeS [34], which has the same structure as SnS.

Based on SE measurements and others such as reflectance [21], transmittance [34], and valence-electron excitation [20], we found that the optical anisotropy of SnS and related materals can be analyzed more precisely by considering the localization of orbitals in the band structure. The CP energies can be determined from the general band structure, while the anisotropic dielectric properties can be explained by the natural anisotropy of the component orbitals. As indicated in previous studies [19,23], the difference between band structures with and without spin-orbit coupling (SOC) is small enough to be negligible in the current analysis of low energy bandgaps, hence non-SOC computation is done here. We verified this in an independent calculation. Figure 3 shows the energy band structure of single crystal SnS. This is plotted with partial orbitals, where blue, magenta, green, and red represent the orbitals $s$, $p_x$, $p_y$, and $p_z$, respectively. The band-to-band transitions corresponding to fundamental band gap $E_0$, $E_1$, and exciton are denoted by arrows in Fig. 3.

As shown in the energy band structure, the top of valance band (VB) is localized along the $\Gamma - Y$ line, while the bottom of conduction band (CB) is at the $\Gamma - X$ line. Therefore, the direct optical transition is not allowed in this situation. Considering the symmetry line $\Gamma - Y$, the indirect band gap is predicted to be the transition from the top of the VB near the $Y-$ point to the first CB at the $\Gamma-$ point, a difference of about 1.10 eV. This value matches well the experimental values of 1.07 [35], 1.05 ± 0.05 [36] and 1.049 eV [37] for the indirect band gap, verifying the validity of current calculation. Unfortunately, SE is not sensitive enough at low absorption coefficients ($\alpha < 100 \text{ cm}^{-1}$) [38], so the indirect fundamental band gap was not observed in Fig. 2.

The fundamental direct band gap ($E_0$) occurs between the first VB and the first CB in the $\Gamma - X$ line. This region is constructed mainly by $s$ and $p_x$ orbitals, as shown by the blue and magenta colors, respectively, in Fig. 3. We note that the minimum position of CB and maximum position of VB do not precisely match. However, the condition for a CP (or van Hove singularity) is that the two bands have the same slope at the same *k*-point. In this case the mismatch is negligible. The calculated result indicates that the energy of this CP is about 1.33 eV. This is in good agreement with the data, which give about 1.39 eV at 27 K. The $E_1$ CP that occurs at 1.66 eV along the armchair-direction is predicted to be the saddle point $M_1$ near the $Y-$ point. The appearance of the exciton at the saddle point is given in Refs. 39 and 40. We find that this feature arises from *s* and $p_y$ orbitals.

An interesting observation seen in Fig. 2 is that most of the CPs have the same peak positions, even though the intensities and lineshapes of the CPs may be much different along the different principal axes. This phenomenon can be interpreted by the natural anisotropy of the *p* orbitals, which leads to the anisotropic localization of wave functions. The contribution of the *d* orbital can be ignored for SnS because Fig. 3 shows its contribution to be small. In Ref. 41 Wang *et al*. show that the intensity of exciton photoluminescence strongly depends on the polarization directions of the source and detector. They found that the wave function of the excitonic feature is significantly extended along the armchair-direction, which results in the highest photoluminescence intensity when the laser and detector polarizations are aligned to the armchair-axis.

To confirm our assumption that the anisotropic properties of SnS comes from the anisotropy of the *p* orbital, we calculate the partial charge density at the top of the VB in the $\Gamma-X$ direction and near the top of the VB in the $\Gamma-Y$ direction, where the $E_0$ and $E_1$ peaks (and its excitonic feature), respectively, are expected. Figure 5(a) shows the partial charge density calculated at the top of the VB in the $\Gamma-X$ region where the $E_0$ peak arises. It is obvious that the charge density is extended along the zigzag-direction, which results in the dominance of the $E_0$ peak in the zigzag-direction, as shown in Fig. 2(b). Similarly, we also observe the preferred orientation of charge density along the armchair-axis at the saddle-point in the $\Gamma-Y$ region where the $E_1$ peak and exciton are detected, as shown in Fig. 2(c,d). This explains the dominance of the exciton along the armchair-axis at low temperatures. These results confirm that the anisotropic dielectric properties of single crystal SnS can be understood fully by considering both energy band structure and the localization of partial orbitals on each energy band.

Figure 5 shows the comparison between the calculated and the measured dielectric functions (blue solid lines) along the three principle axes of SnS. The black dashed-dotted curves are the results of the DFT calculations that do not take the electron-hole interaction into account. It is obvious that the DFT calculations are not well matched to the data, as are the BSE calculations which consider the interaction of electron-hole as shown by red dashed curves. In particular, the excitonic peak is clearly predicted in the BSE calculation in the armchair-direction, as shown in Fig. 5(d). The good agreement between the current calculation and reported SE spectra confirms that the accuracy of our theoretical approach is acceptable.

## IV. CONCLUSIONS

Based on the dielectric tensor of single-crystal SnS measured at low temperature (27 K) and *ab initio* calculations, we report an approach for investigating the anisotropic properties of single-crystal SnS. The combination of band-structure calculations and partial orbitals allows a more insightful analysis of the physical meaning of the anisotropic properties of single-crystal SnS. From the SE data, the excitonic feature strongly dominates the armchair direction. This is understood by the band-to-band transitions at the saddle point in $\Gamma - Y$ region, where the partial charge density has its major orientation along to the armchair-axis.

## ACKNOWLEDGMENTS

We acknowledge S. Cho in University of Ulsan for providing samples. This research was supported by a National Research Foundation of Korea (NRF) grant funded by the Korea government (MSIP) (NRF-2020R1A2C1009041) and the National Research Foundation of Korea (NRF) grant funded by the Korea government (NRF-2019R1H1A2079786).

**FIGURE CAPTIONS**

FIG. 1. (a) The distorted-rocksalt orthorhombic (Pnma) crystal structure of SnS. The Sn atoms are dark grey and the S atoms are bright grey; (b) Brillouin zone of the SnS structure. (c) The change of dielectric function from the zigzag-direction to the armchair-direction, illustrated by changing the azimuthal angle (room-temperature data). (d) Mapping of the full azimuthal rotation angle of the dielectric function on the cleavage plane of single crystal SnS.

FIG. 2. Real and imaginary parts of the dielectric tensor of SnS in the cleavage plane at (a-b) 27 K and (c-d) 300 K. Dashed, solid, and dashed-dotted curves show the dielectric tensor at azimuthal angles of $0°$ ($a$-axis), $45°$, and $90°$ ($b$-axis), respectively.

FIG. 3. Band structure of SnS with partial orbitals of atoms are presented for each energy band. Blue, magenta, green, and red represent $s$, $p_x$, $p_y$, and $p_z$, respectively.

FIG. 4. Partial orbitals (partial charge density) (a,b) at the top of valance band in the $\Gamma - X$ line and (c,d) at saddle point near the top of valance band in the $\Gamma - Y$ line. The Sn atoms are dark grey while the S atoms are bright grey.

FIG. 5. Calculated dielectric functions using DFT and BSE compared with corrected- data along three principal axes. Details are given in the text.

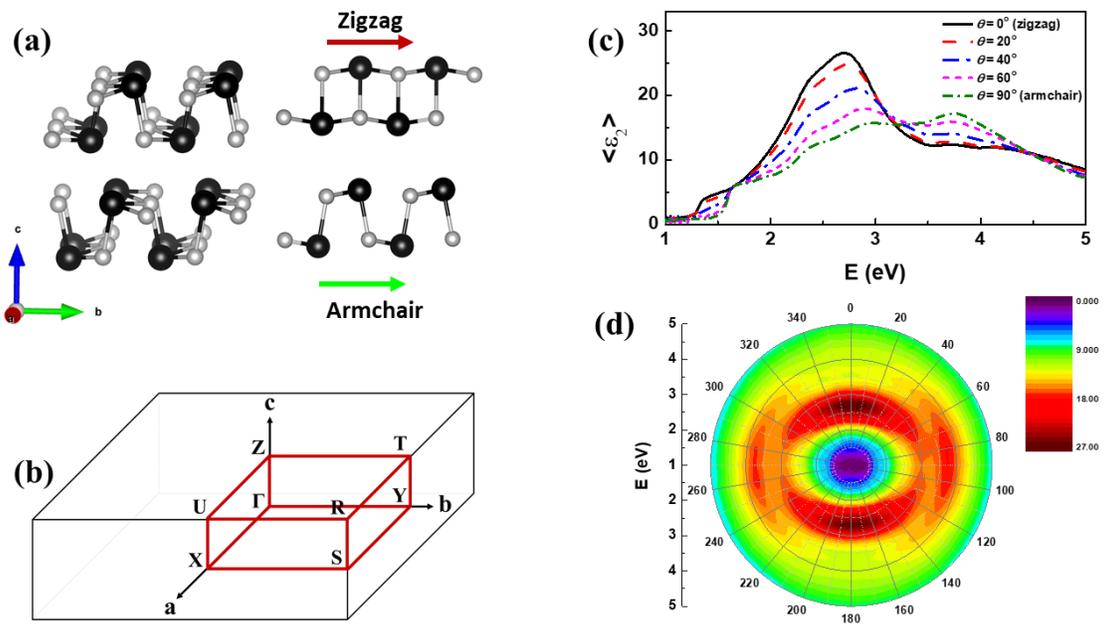

**FIG. 1.**

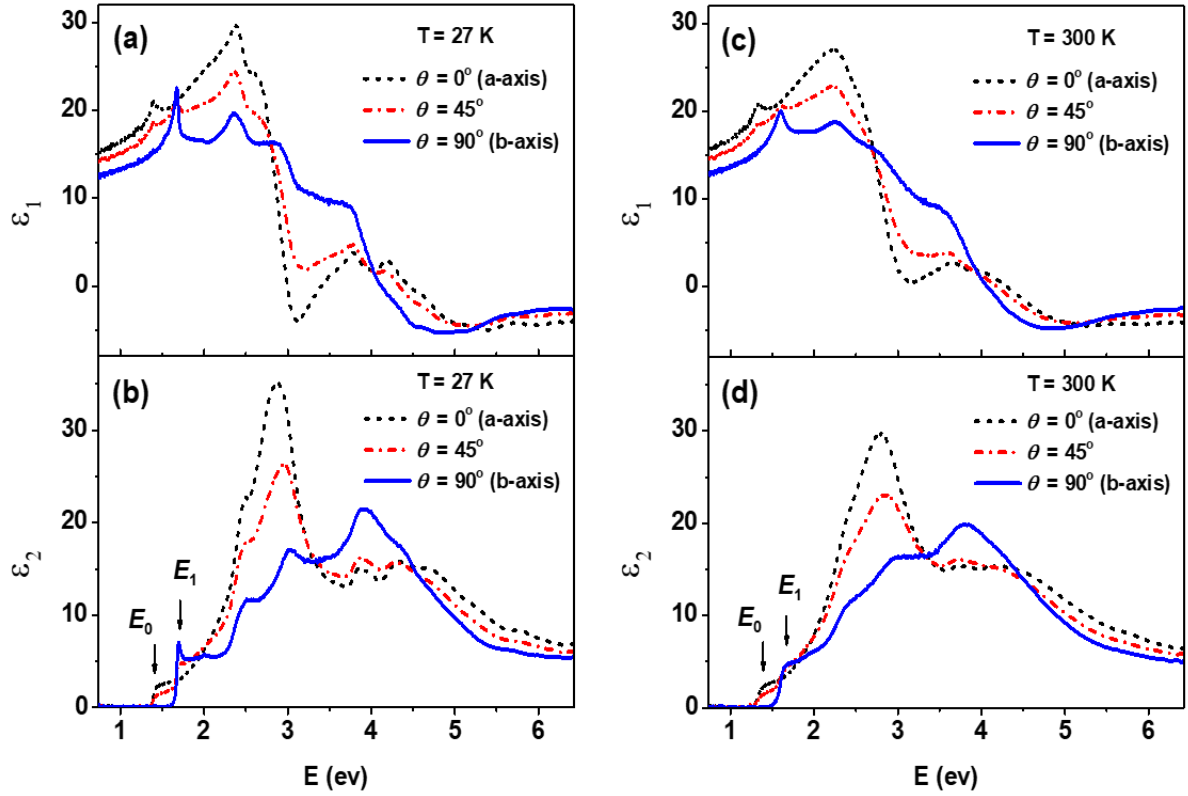

**FIG. 2.**

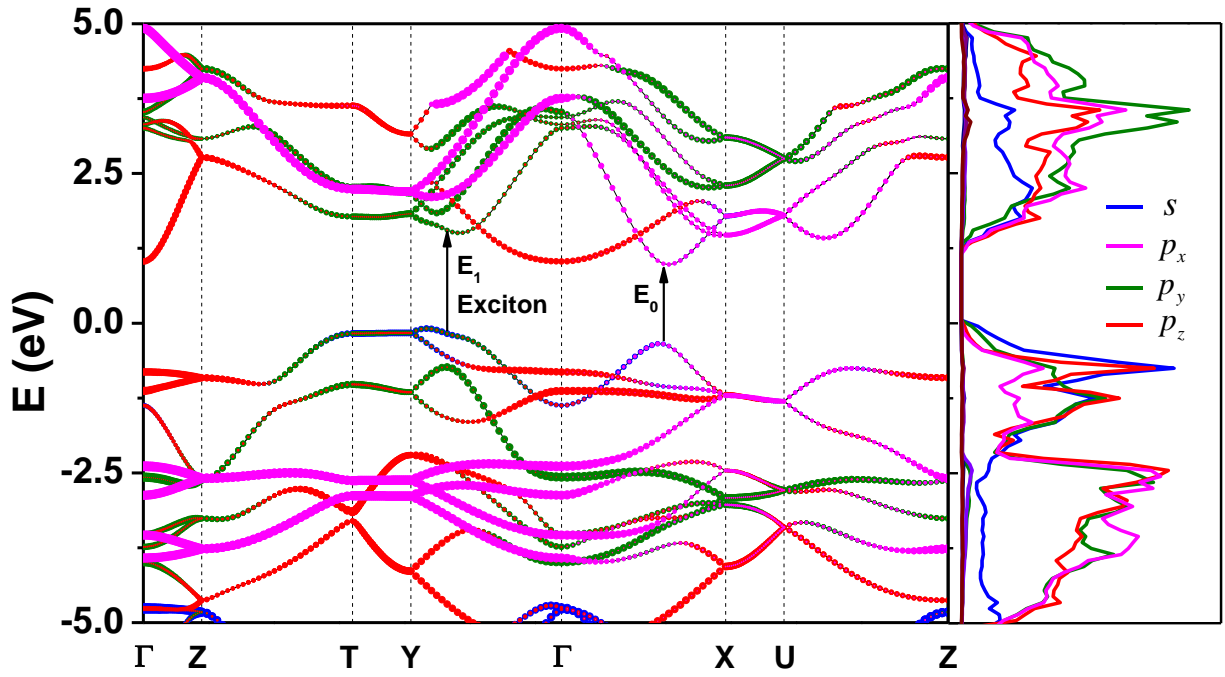

**FIG. 3.**

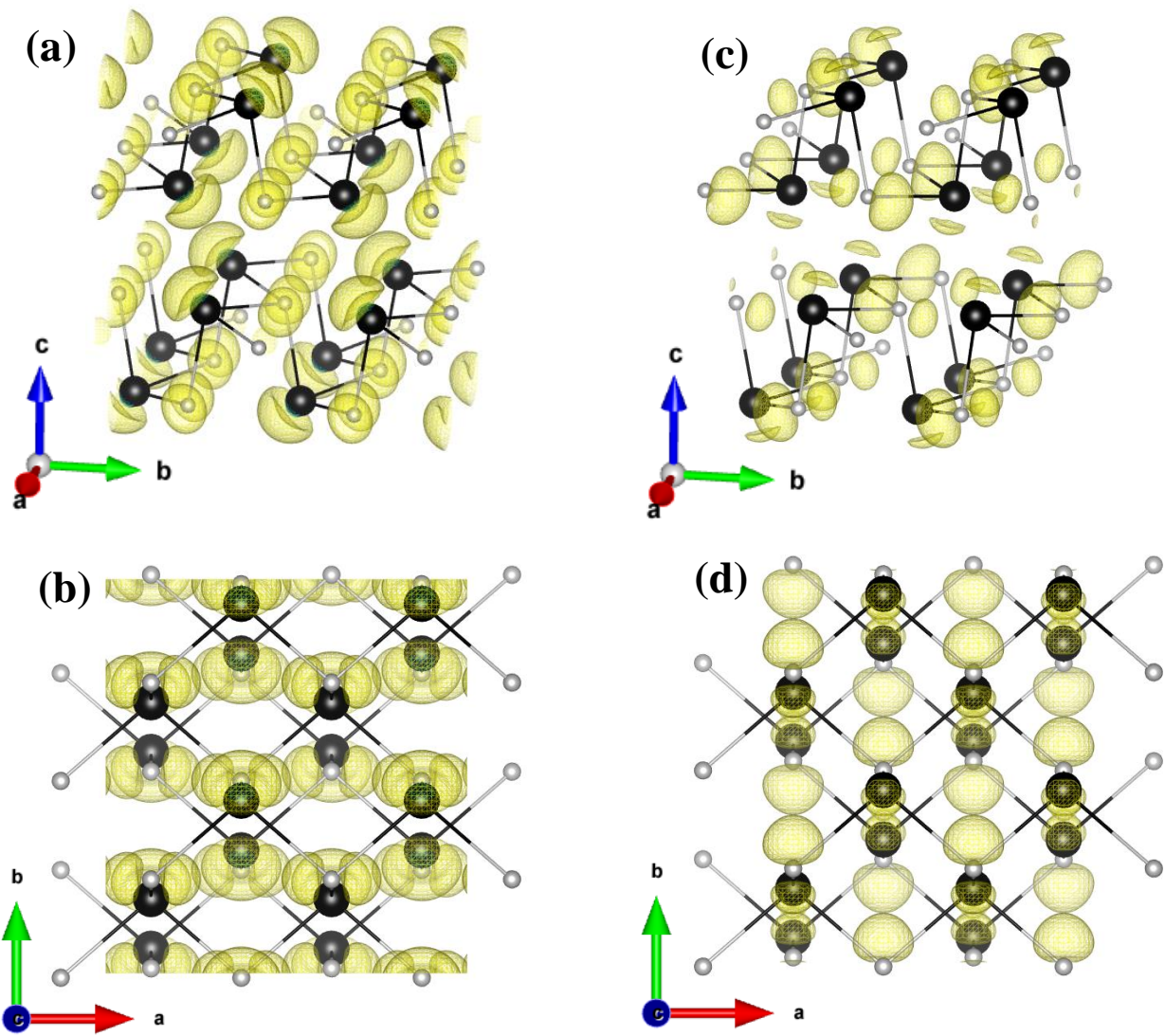

**FIG. 4.**

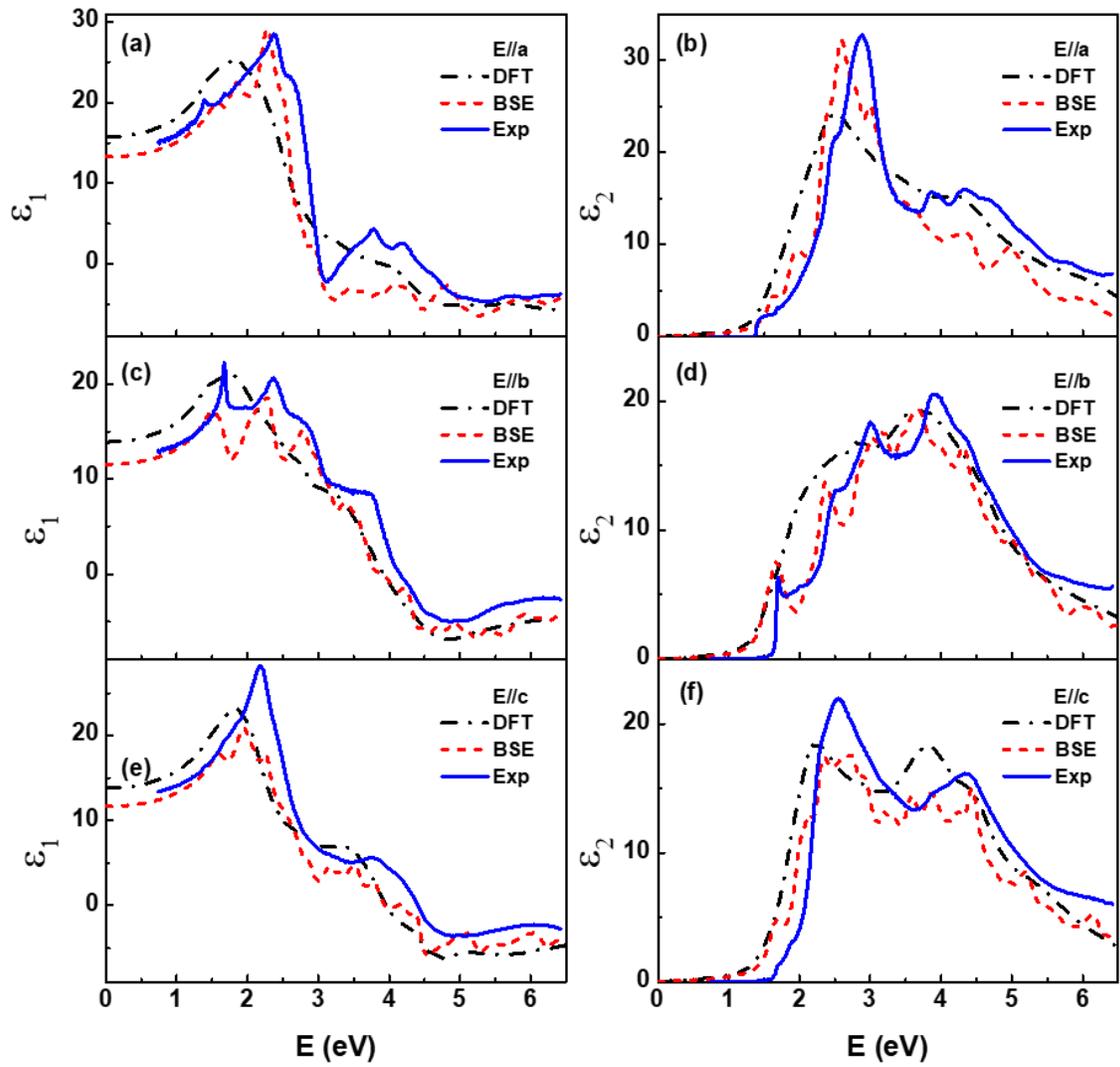

**FIG. 5.**